\documentstyle[12pt]{article}

\newcommand{\beq}{\begin{equation}}
\newcommand{\eeq}{\end{equation}}
\newcommand{\bea}{\begin{eqnarray}}
\newcommand{\eea}{\end{eqnarray}}

\newcommand{\gsim}{\lower.7ex\hbox{$
\;\stackrel{\textstyle>}{\sim}\;$}}
\newcommand{\lsim}{\lower.7ex\hbox{$
\;\stackrel{\textstyle<}{\sim}\;$}}

\newcommand{\La}{\overline{\Lambda}}

\newcommand{\eod}{\end{document}}

\begin{document}
\thispagestyle{empty}
\vspace*{-20mm}

\begin{flushright}
UND-HEP-05-BIG\hspace*{.08em}07\\
LPT-ORSAY 05-85\\
LAL 05-178\\
hep-ph/0512270\\
%version 3.0\\

%\today 

\end{flushright}
\vspace*{7mm}

\begin{center}
{\LARGE{\bf
Memorino on the `1/2 vs. 3/2 Puzzle' in 
$\bar B \to l \bar \nu X_c$}}
\vspace*{14mm}

{\large{\bf I.I.~Bigi$^{\,a}$, B. Blossier$^c$, A. Le Yaouanc$^c$, 
L. Oliver$^c$, O. P\`ene$^c$, J.-C. Raynal$^c$,  
A. Oyanguren$^d$, P. Roudeau$^d$}}\\
\vspace{4mm}

$^a$ {\sl Department of Physics, University of Notre Dame du Lac}
\vspace*{-.8mm}\\
{\sl Notre Dame, IN 46556, USA}\vspace*{1.5mm}\\ 
$^c$ {\sl LPTh, Univ. de Paris Sud, F-91405 Orsay CEDEX, France} \\ 
$^d$ {\sl LAL, F-91898 Orsay CEDEX, France}

\vspace*{5mm}

{\bf Abstract}\vspace*{-.9mm}\\
\end{center}

\noindent
After the successes the OPE description has scored in describing $\bar B \to l \bar \nu X_c$ decays, we 
need to study what can be said about the composition of the hadronic final state $X_c$. 
The same OPE treatment yields sum rules implying the dominance of $j_q = 3/2$ 
charm states in $X_c$ over their $j_q=1/2$ counterparts. This prediction is supported by other 
general arguments as well as quark model calculations. At present 
it is unclear to which  degree data  conform to these predictions. More experimental information is essential. We want to ask our experimental colleagues for a redoubled effort to establish, which 
hadronic configurations -- $D/D^* + \pi, D/D^* + 2 \pi, ...$ -- make up 
$\Gamma (\bar B \to l \bar \nu X_c)$ 
beyond $\bar B \to l \bar \nu D/D^*$, what their quantum numbers are and their mass distributions. 
The latter is most relevant for the determination of hadronic mass moments in 
$\bar B \to l \bar \nu X_c$. Since all this will require considerable effort on their part, we want to explain the 
theoretical issues involved, why they carry `gravitas' -- i.e. are weighty -- and why a better understanding of them will be of significant value. In this brief memo we sketch the underlying 
arguments based on heavy quark theory, the OPE, a special class of quark models and lattice 
QCD in a nutshell. After summarizing the experimental situation we conclude with two lists, namely one 
with measurements that need to be done and one with items of theoretical homework. Some of the 
latter can be done by employing existing theoretical tools, whereas others need new ideas.

\vspace*{10mm}
%%%%%%%%%
\section{Outline}
%%%%%%%%

This brief memo follows from discussions between theorists and experimentalists on the 
composition of semileptonic $B$ decays {\em beyond} $\bar B \to l \bar \nu D/D^*$. 
Understanding the nature of the 
hadronic system in the final state -- its quantum numbers as well as mass distributions -- is 
important, since well grounded theoretical expectations and predictions can and have been 
given on these issues. In particular our theoretical understanding of semileptonic $B$ decays 
tells us that hadronic charm states where the light degrees of freedom carry angular momentum 
$j_q = 3/2$ should be considerably more abundant in the final states than their $j_q = 1/2$ 
counterparts.  This prediction appears to be at variance with some data. 
We refer to this apparent conflict as the `1/2 vs. 3/2 puzzle' \cite{PUZZLE}. The aim of 
this note is to explain in a concise way what is involved in the argument to stimulate further studies. 

We marshall the theoretical arsenal for treating semileptonic $B$ decays into charm hadrons: the operator product expansion (OPE) in Sect.{\ref{OPETR}, the BT model in 
Sect.\ref{BTMOD} and lattice QCD in Sect.\ref{LQCD} before adding other general arguments in 
Sect.\ref{GENERAL}; in Sect.\ref{EXPSTAT} we comment on the existing data on 
$\bar B \to l \bar \nu D^{(*)} + \pi$'s from ALEPH, BELLE, CDF, DELPHI and D0 before listing 
needed homework for both theorists and experimentalists in Sect.\ref{ACTION}. We aim at being as concise as reasonably possible, while providing a guide through the literature for the more committed reader. 

%%%%%%%%%%%
\section{The OPE treatment}
\label{OPETR}
%%%%%%%%%%
Both our theoretical and experimental knowledge on semileptonic $B$ decays have advanced 
considerably over the last 15 years. This progress can be illustrated most strikingly by the recent success in extracting the value of $|V(cb)|$ with better than 2 \% accuracy from measurements of 
$\bar B \to l \bar \nu X_c$ transitions \cite{FLAECHER}. At the same time it has also brought various 
potential problems into 
sharper focus. One concerns the size of $BR_{SL}(B)$. However in this memo we want to focus 
on the composition of the hadronic final state in $\bar B \to l \bar \nu X_c$. 

While the OPE allows to describe inclusive transitions, no 
{\em systematic} extension to exclusive modes has been given so far. Yet even so, the OPE allows to place important constraints on some exclusive rates: 
$\bar B\to l \bar \nu D^*$, $l \bar \nu D$ (the latter involving the `BPS' expansion) are the most topical and elaborated 
examples \cite{HQE,BPS}. 

OPE results can be given also for subclasses of inclusive transitions due to various  
sum rules \cite{URSR,ORSAYSR} that can genuinely be derived from QCD; hence one can infer constraints on certain exclusive contributions.   
Those can be formulated most concisely when one adopts the heavy quark symmetry classification scheme also for the hadronic charm system in the final state of semileptonic $B$ decays. 

In the limit $m_Q \to \infty$ one has heavy quark symmetry controlling the spectroscopy for mesons as follows: The heavy quark spin decouples from the dynamics, and the hadrons can be labeled by their total spin $S$ together with the angular momentum $j_q$ carried by the light degrees of freedom, namely the 
light quarks and the gluons. The pseudoscalar and vector mesons 
$D$ and $D^*$ then form the ground states of heavy quark symmetry in the charm sector with 
$j_q=1/2$. The first excited states are four P wave configurations, namely two with $j_q = 3/2$ and 
$S=2,1$ -- $D^{3/2}_2$, $D^{3/2}_1$ -- and two with $j_q = 1/2$ and $S=1,0$ -- 
$D^{1/2}_1$, $D^{1/2}_0$; the two $3/2$ states are narrow resonances and the 
two $1/2$ states wide ones.  Then there are higher states still, namely radial excitations and higher 
orbital states; furthermore there are charm final states that cannot be properly called a hadronic resonance, but are $D/D^*$ combinations with any number of pions etc. carrying any 
allowed $J^{PC}$ quantum numbers. 

The usual Isgur-Wise function $\xi (w)$ is the core element in describing $\bar B\to l \bar \nu D/D^*$. It can 
be generalized to describe also the production of excited charm final states in semileptonic $B$ decays: $\tau _{1/2[3/2]}^{(n)}(w_n)$ with $w_n = v_B\cdot v_{D^{(n)}}$ is the amplitude for 
$\bar B\to l \bar \nu D^{(n)}_{1/2[3/2]}$, 
where $D^{(n)}_{1/2[3/2]}$ denotes a hadronic system with open charm carrying 
$j_q = 1/2[3/2]$ and label $n$; it does {\em not} need to be a bona fide 
resonance. 

Various sum rules can be derived from QCD proper relating the moduli of these amplitudes and 
powers of the excitation energies $\epsilon _n = M_{D^{(n)}} - M_D$ to heavy quark parameters. 
Adopting the so-called `kinetic scheme', as we will throughout this memo, one obtains in particular 
\cite{URSR}: 
\bea
\frac{1}{4} &=& - \sum _n \left| \tau_{1/2}^{(n)}(1)\right| ^2 + 
\sum _m \left| \tau_{3/2}^{(m)}(1)\right| ^2 
\label{SR1}
\\
\mu_{\pi}^2 (\mu)/3 &=& \sum _n^{\mu} \epsilon_n^2\left| \tau_{1/2}^{(n)}(1)\right| ^2 +
2\sum _m^{\mu} \epsilon_m^2\left| \tau_{3/2}^{(m)}(1)\right| ^2 
\label{SR2}
\\
\mu_{G}^2 (\mu)/3 &=& 
-2\sum _n^{\mu} \epsilon_n^2\left| \tau_{1/2}^{(n)}(1)\right| ^2 +
2\sum _m^{\mu} \epsilon_m^2\left| \tau_{3/2}^{(m)}(1)\right| ^2
\label{SR3}
\eea
where the summations go over all hadronic systems with excitation energies 
$\epsilon_{n,m} \leq \mu$ 
\footnote{The sum rule of Eq.(\ref{SR1}) does not require a cut-off or normalization scale 
$\mu$, as is already implied by its left hand side \cite{URSR}.}. 
The sum rules show that the heavy quark parameters 
$\mu_{\pi}^2$ and $\mu_G^2$ (and likewise for $m_b$, $m_c$) defined in the 
kinetic scheme are {\em observables}. 

These sum rules allow us to make both general qualitative as well as 
(semi)quantitative statements. 
On the {\em qualitative} level we learn unequivocally that the `3/2' transitions have to dominate over the 
`1/2' ones, as can be read off from Eqs.(\ref{SR1},\ref{SR3}). 
Furthermore we know that $\mu_{\pi}^2 (\mu ) \geq \mu _G^2(\mu )$ has to hold for any $\mu$ 
\cite{OPTICAL}, as is actually obvious from Eqs.(\ref{SR2},\ref{SR3}).  
On 
the {\em quantitative} level  it is not a priori clear, at which scale 
$\mu$ these sum rules are saturated and by which kind of states. 

To address those issues we can invoke some rules of thumb (not to be confused 
with sum rules) gleaned from previous experience. That tells us that the sum rules should 
saturate to a decent degree of accuracy at $\mu \sim 1$ GeV 
through the four P wave states. 

We have learnt a lot about the numerical values of the heavy quark parameters: the most accurate value for the chromomagnetic moment $\mu_G^2$ can be deduced from the $B^*-B$ hyperfine mass splitting: 
\beq 
\mu_G^2(1 \; {\rm GeV}) = (0.35 \pm 0.03) \; {\rm GeV}^2
\eeq
A comprehensive study of energy and hadronic mass moments in $\bar B \to l \bar \nu X_c$ 
yields \cite{FLAECHER}: 
\beq 
\mu_{\pi}^2 (1 \; {\rm GeV}) = (0.401 \pm 0.04) \; {\rm GeV}^2
\eeq
These values for 
$\mu_{\pi}^2 (1\; {\rm GeV})$ and $\mu_{G}^2 (1\; {\rm GeV})$ show that 
$\mu_{\pi}^2 (1\; {\rm GeV}) - \mu_{G}^2 (1\; {\rm GeV}) \ll
\mu_{\pi}^2 (1\; {\rm GeV}) + \mu_{G}^2 (1\; {\rm GeV})$ holds, which 
forms one of the cornerstones of the BPS expansion. Inserting these values into 
\beq 
\mu_{\pi}^2 (\mu) - \mu_{G}^2 (\mu) = 
9 \sum _n \epsilon_n^2\left| \tau_{1/2}^{(n)}(1)\right| ^2 \; , 
\eeq 
see Eqs.(\ref{SR2},\ref{SR3}), shows the `1/2' states contribute very little both to the sum rules 
and to semileptonic $B$ decays. Moreover an upper bound can be placed on the 
production of the lowest `1/2' states: 
\beq 
\frac{1}{9}[\mu_{\pi}^2 (\mu) - \mu_{G}^2 (\mu)] \geq 
\epsilon_0^2\left| \tau_{1/2}^{(0)}(1)\right| ^2 \; . 
\eeq

The data tell us that $\bar B\to l \bar \nu D/D^*$ make up about three quarters of the inclusive semileptonic 
$B$ width \cite{ARAN}: 
\bea 
{\rm BR}(\bar B_d \to l \bar \nu X_c) &=& (10.31 \pm 0.15)\% 
\label{SLBR}
\\
{\rm BR}(\bar B_d \to l \bar \nu X_c) - {\rm BR}(\bar B_d \to l \bar \nu D)  - {\rm BR}(\bar B_d \to l \bar \nu D^*) 
&=& (2.9 \pm 0.3)\%
\label{SLBRD**}
\eea
The dominance of these two final states represents actually the most direct evidence that charm quarks act basically like heavy quarks in $B$ decays. This can be invoked to justify the use of the heavy quark 
classification already to charm. 

A natural `scenario' for the implementation of the OPE description and its sum rules is provided by 
\bea 
 |\tau _{3/2}^{(0)}(1)|^2 &\simeq& 0.3 \; , \; \; \; \; \; \; 
\epsilon_{3/2} \sim 450 \; {\rm MeV} \\
|\tau _{1/2}^{(0)}(1)|^2 &\simeq& 0.07 - 0.12 \; , \; 
\epsilon_{1/2} \sim (300 - 500) \; {\rm MeV} 
\eea
Finally there is no reason why the six final states $D$, $D^*$, $D^{3/2}_{2,1}$ and 
$D^{1/2}_{1,0}$ should saturate 
$\Gamma _{SL}(B)$. One actually expects QCD radiative corrections to populate the higher 
hadronic mass region above the prominent resonances through a smooth spectrum dual to a 
superposition of broad resonances. 

%%%%%%%%%%
\section{The BT model}
\label{BTMOD}
%%%%%%%%

Based on the OPE treatment alone one cannot be more specific numerically. 
To go further one relies on quark models for guidance. The dominance of the `3/2' over the `1/2' states emerges naturally in 
all quark models obeying known constraints from QCD as well as Lorentz covariance. 
This can be demonstrated  explicitly with the Bakamjian-Thomas covariant quark model  
\cite{BTMOD}, which satisfies heavy quark symmetry and the Bjorken as well as spin sum rules referred to above. It allows 
to determine the masses of various charm excitations and 
to compute the production rates in semileptonic  \cite{BTORSAY1,BTORSAY2,HYCHENG} 
as well as nonleptonic $B$ decays \cite{BTORSAY2}. 
The BT model provides a quantitative illustration of the heavy quark limit, in particular concerning 
the sum rule of Eq.(\ref{SR1}). One finds 
\bea 
\tau ^{(0)}_{1/2}(1) &=& 0.22  \\
\tau ^{(0)}_{3/2}(1) &=& 0.54  
\eea
together with predictions for the slopes. 
For the semileptonic modes the BT model yields: 
\bea 
{\rm BR}(\bar B \to l \bar \nu D) &=& (1.95 \pm 0.45) \% \; , \; 
\label{BRTH1}
\\
{\rm BR}(\bar B \to l \bar \nu D^*) &=& (5.90 \pm 1.10) \% 
\label{BRTH2}
\\
{\rm BR}(\bar B \to l \bar \nu D_2^{3/2}) &=& (0.63^{+0.3}_{-0.2}) \% 
\label{BRTH3} 
\\
{\rm BR}(\bar B \to l \bar \nu D_1^{3/2} )&=& (0.40^{+0.12}_{-0.14}) \%
\label{BRTH4}
\\
{\rm BR}(\bar B \to l \bar \nu D_1^{1/2}) &=& (0.06 \pm 0.02) \% 
\label{BRTH5} 
\\
{\rm BR}(\bar B \to l \bar \nu D_0^{1/2}) &=& (0.06 \pm 0.02) \%
\label{BRTH6}
\eea
The following features of the model predictions should be noted in particular: they 
\begin{itemize}
\item 
agree with the data on $\bar B\to l \bar \nu D/D^*$,  
\item 
exhibit a strong dominance of `3/2' over `1/2' production as inferred already from 
the Sum Rules and   
\item 
appear to fall somewhat short of saturating the observed $\Gamma_{SL}(B)$; 
this last feature is to be expected on general grounds as indicated at the end of the previous section. 
\item
Last, but not least (although it is not the focus of this memo), the model provides a nice description 
of the nonleptonic modes $\bar B \to D/D^* \pi$. For our purposes this is particularly 
significant in the channel 
$\bar B \to D^{**+}\pi^-$ measured recently \cite{BELLENL}, where factorization can be justified 
\cite{BTORSAY2};  the rate thus provides direct information on $\tau ^{(0)}_{1/2}(w)$ 
and $\tau ^{(0)}_{3/2}(w)$. 

\end{itemize}

%%%%%%%%
\section{Lattice QCD}
\label{LQCD}
%%%%%%%%%

In principle the two form factors $\tau_{1/2}(1)$ and $\tau_{3/2}(1)$ can be computed in a straightforward way using the HQET equation of motion $(v\cdot D) \,h_v=0$ \cite{lig}:
\bea\nonumber
_v\langle 0^+| \bar{h}_v \gamma^i \gamma^5 D^j h_v|0^-\rangle_v &=& i\, g^{ij}\,
\tau_{1/2}(1)\, (\La_{0^+}-\La_{0^-}),\\
_v\langle 2^+| \bar{h}_v \left(\frac{\gamma^i \gamma^5 D^j+\gamma^j \gamma^5 D^i}{2}\right) 
h_v|0^-\rangle_v &=& -i\sqrt{3}\,\epsilon^{*ij}\, \tau_{3/2}(1)\, (\La_{2^+}-\La_{0^-}),
\eea
where $v=(1,\vec{0})$ is the heavy quark velocity, $\epsilon^*$ the 
polarization tensor of the $2^+$ state and
$\La_{\rm J^P}$ the dominant term in the OPE expression for the ${\rm J^P}$ heavy-light meson  
binding energy. On 
the lattice the covariant derivative $D_i$ applied to the static quark field $h(\vec{x},t)$ is 
expressed as 
$D_i\, h (\vec x,t)\to \frac 1{2a}\left (U_i(\vec x,t)h (\vec x+\hat i,t) 
- U_i^\dagger(\vec x-\hat i,t) h (\vec x-\hat i,t) \right )$; $U_i(\vec x,t)$ denotes the gauge link. 
One calculates as usual the two-point functions $C^2_{\rm J^P}(t) =
\langle 0|O_{\rm J^P}(t)O_{\rm J^P}^\dag(0)|0\rangle$,  the three-point functions 
$C^3_{\rm J^P,0^-}(t_1,t_2)=\langle 0|O_{\rm J^P}(t_2)O_\Gamma(t_1)O_{0^-}^\dag(0)|0\rangle$ 
and  
$\langle J^P|O_\Gamma|0^-\rangle \propto 
R(t_1,t_2)=\frac{C^3_{\rm J^P,0^-}(t_1,t_2)}{C^2_{0^-}(t_1)C^2_{\rm J^P}(t_2-t_1)}$.

Alas, numerical complications appear, because orbital as well as radial excitations can contribute. 
To extract properly the matrix element for the P wave state 
$\langle J^P|O_\Gamma|0^-\rangle$, one has to effectively suppress 
the coupling of radial excitations (with quantum numbers $n>1,\, J^P$) 
to the vacuum. This can be achieved by choosing an appropriate interpolating field 
$O_{\rm J^P}$ such that $\langle n>1\, J^P|O_{\rm J^P}|0 \rangle=0$ holds or by having 
huge statistics to diminish statistical fluctuations at large times (where 
the fundamental state is no more contaminated by radial excitations). 
This poses a problem in particular for the $2^+$ state, for which the usual interpolating 
field seems to couple also the  first radial excitation quite strongly to the vacuum. Moreover 
reaching the required stability of $R(t_1,t_2)$ as a function of $t_2$ poses a serious 
challenge. 

Therefore we will need very careful and dedicated lattice studies to obtain meaningful and reliable 
results for $\tau_{3/2,1/2}$. As an already highly relevant intermediate step 
one can concentrate first on $\tau_{1/2}$ to see whether lattice QCD confirms its suppression 
as inferred from both the sum rules and the BT model. A preliminary study in the quenched approximation with $\beta=6.0\, (a^{-1}=2 \,{\rm GeV^{-1}})$ and $m_q \simeq m_s$ yields 
\cite{tauorsay}: 
\bea 
\tau_{1/2}^{(0)}(1) &\sim& 0.3\div 0.4 \\
\tau_{3/2}^{(0)}(1) &\sim& 0.5\div 0.6 \; . 
\eea
Apart from unquenching and lowering the value of $m_q$ 
one can improve and refine this analysis also by simulating a {\em dynamical} charm quark, 
i.e. applying HQET to the $B$ meson only. This would allow to evaluate $1/m_c$ corrections. 

%%%%%%%%%%%%%
\section{Two other general arguments on $|\tau_{1/2}/\tau_{3/2}|^2$}
\label{GENERAL}
%%%%%%%%%

The numerics of the theoretical predictions on semileptonic $B$ decays given above have to be taken `cum grano salis'. Yet their principal feature -- the preponderance of `3/2' over `1/2' states -- has to be taken very seriously, since it is supported  by two rather general observations that point in the same direction as the detailed theoretical considerations given above:  
\begin{itemize}
\item 
When interpreting data one should keep in mind that the contributions of $D^{1/2}_{1,0}$ to  
$\Gamma (\bar B \to l \bar \nu D^{**})$ 
\footnote{$D^{**}$ will be used as a short-hand for $D/D^* + \pi$'s final states.} 
are suppressed relative to those from  $D^{3/2}_{2,1}$ by a factor of two to three 
due to kinematics \cite{BTORSAY1}. 
Thus one finds for reasonable values of $\tau^{(0)}_{1/2}$ that 
$\Gamma(\bar B \to l \bar \nu D^{1/2})$ 
falls below $\Gamma (\bar B \to l \bar \nu D^{3/2})$ by one order of magnitude, as illustrated below. 
For the two widths to become comparable, one would need a greatly enhanced 
$\tau^{(0)}_{1/2}$. 
\item 
There is a whole body of evidence showing that in so-called class I nonleptonic $B$ decays like 
$\bar B_d \to D^{(*)+}\pi^-$ naive factorization provides a very decent description of the data. Invoking 
this ansatz also for $\bar B_d \to D^{**+}\pi^-\to D^{(*)0} \pi ^+\pi^-$ one infers from BELLE's data 
\cite{BELLENL} that the production of `1/2' states appears to be strongly suppressed relative to that 
for `3/2' ones. It implies that  $|\tau_{1/2}/\tau_{3/2}|^2$ is small and certainly less than unity. 
This agrees with the theoretical expectations described before; more importantly it shows 
in a rather model independent way that there is no large unexpected enhancement of  
$|\tau_{1/2}|$. Those values also allow to saturate the sum rule of Eq.(\ref{SR1}) 
within errors already with the $n=0$ states. 

The form factors are actually probed at $w=1.3$ in this nonleptonic transition; yet a natural functional  dependence on $w$ supports this conclusion to hold for $1 \leq w \leq 1.3$ in semileptonic 
channels. 
\end{itemize}
These arguments are based on the heavy quark mass limit. The as yet unknown finite mass corrections 
could modify these conclusions somewhat. 

%%%%%%%%%%
\section{Comparison with the data on semileptonic $B$ decays}
\label{EXPSTAT}
%%%%%%%%

The measurements on the production of the `3/2' states are consistent and agree with the 
theoretical expectations of a total branching ratio of ${\cal O}(1\% )$. The disagreements concern the production of the `1/2' states as 
well as radial and higher orbital excitations as explained below.

ALEPH \cite{ALEPHEXP} has reconstructed $D^{**}$ states decaying into 
$D^{(*)}\pi^{\pm}$. They did not observe a significant excess of events over the expected background 
in $D^{(*)+}\pi^+$ or $D^0\pi ^-$ combinations (called `wrong sign'). From the measured rate of `right sign' combinations and assuming that only $D^{**}$ decaying to $D^{(*)}\pi$ contribute (to correct for 
channels with a missing $\pi^0$) they get (with Prob$(b \to B)=(39.7 \pm 1.0)\%$): 
\beq 
{\rm BR}(\bar B \to l \bar \nu D^{**}) = (2.2 \pm 0.3 \pm 0.3) \% 
\label{ALEPHBR1}
\eeq
Assuming the $D^{3/2}_1$ state to decay only into $D^*\pi$ they find also:
\bea 
{\rm BR}(\bar B \to l \bar \nu D^{3/2}_1) &=& (0.70 \pm 0.15) \% \\
{\rm BR}(\bar B \to l \bar \nu D^{3/2}_2) &<& 0.2 \% \; , 
\label{ALEPHBR2}
\eea
numbers which are not in conflict with the BT predictions, Eqs.(\ref{BRTH3}, \ref{BRTH4}). 
From their observed number of `wrong sign' combinations one can infer (90\% C.L.)
\beq 
{\rm BR}(\bar B \to l \bar \nu D^{*}\pi \pi) \leq 0.35 \% \; , \; {\rm BR}(\bar B \to l \bar \nu D\pi \pi) \leq 0.9 \%
\label{ALEPHBR3}
\eeq

DELPHI has published a re-analysis of their data \cite{DELPHI} superseding their previous 
study \cite{DELPHIPREVIOUS}. Like ALEPH they have not found evidence for 
$\bar B \to l \bar \nu D^{**}\to l \bar \nu D^{(*)}\pi \pi$; assuming only $D^{(*)}\pi$ to contribute 
(to correct for channels with a missing $\pi^0$), they obtain: 
\beq 
{\rm BR}(\bar B_d \to l \bar \nu D^{**}) = (2.7 \pm 0.7 \pm 0.2) \% 
\eeq
This value is sufficient to saturate $\Gamma _{SL}(B)$, see Eq.(\ref{SLBRD**}). 
DELPHI has obtained clear evidence for two narrow states tentatively identified with $D^{3/2}$ 
\beq 
{\rm BR}(\bar B \to l \bar \nu D_1^{3/2} )= (0.56 \pm 0.10)  \% \; , \; 
{\rm BR}(\bar B \to l \bar \nu D_2^{3/2}) = (0.30 \pm 0.08)
\label{DELPHI3/2} \; , 
\eeq 
again in rough agreement with Eqs.(\ref{BRTH3}, \ref{BRTH4}). 

However they found a significant rate for producing a broad hadronic mass distribution: 
\beq 
{\rm BR}(\bar B \to l \bar \nu D_{"1"}) = (1.24 \pm 0.25 \pm 0.27) \% \; , \; 
{\rm BR}(\bar B \to l \bar \nu D_{"0"})= (0.65 \pm 0.69) \% \;  ,  
\label{DELPHI1/2}
\eeq
which appears to be in conflict with the predictions of Eqs.(\ref{BRTH5}, \ref{BRTH6}). 

From the analysis of `wrong sign' combinations they infer the following limits
\beq 
{\rm BR}(\bar B \to l \bar \nu D^{*}\pi \pi) \leq 1.2 \% \; , \; {\rm BR}(\bar B \to l \bar \nu D\pi \pi) \leq 1.3 \%
\label{DELPHIBRPIPI}
\eeq
Considering that $1^+$ $D^{**}$ can decay into $D\pi \pi$ and analyzing the $D\pi$ mass 
distribution they fit a value of $(19\pm 13)\%$ for this component. In their analysis of hadronic 
mass moments such a possibility has been included with 
BR$(\bar B \to l \bar \nu D\pi\pi) = (0.36 \pm 0.27)\%$. This turns out to be the dominant systematic uncertainty 
in their hadronic mass moment measurement. 

The D0 collaboration has measured production rates of narrow $D^{**}$  states in the decay 
$\bar B \to \mu ^- \bar \nu D^*\pi$. Assuming BR$(D_1^{3/2}\to D^*\pi)=100\%$ and 
BR$(D_2^{3/2}\to D^*\pi)=(30 \pm 6)\%$ they obtain \cite{DZERO}
\beq
{\rm BR}(\bar B \to \mu ^- \bar \nu D_1^{3/2}) = (0.33 \pm 0.06) \% \; , \; 
{\rm BR}(\bar B \to \mu^- \bar \nu D_2^{3/2})= (0.44 \pm 0.16) \% \;  . 
\label{D0DATA}
\eeq

{\em If} the broad contributions were indeed to be identified with the $D_{1,0}^{1/2}$ as 
already implied in Eq.(\ref{DELPHI1/2}) -- an a priori reasonable working 
hypothesis -- one would have a clear cut and significant conflict with the OPE expectations 
as well as the numerically more specific BT model predictions. For DELPHI's data would yield  
$\Gamma (\bar B \to l \bar \nu D^{1/2}) > \Gamma (\bar B \to l \bar \nu D^{3/2})$. This conflict has been referred to 
as the `1/2 $>$ 3/2 puzzle' \cite{PUZZLE}. Since, as sketched before, the theoretical predictions are based on a 
rather solid foundation, they should not be discarded easily.  Of course there is no proof that the broad $D/D^*+\pi$ systems are indeed the 
$j_q = 1/2$ P wave states; they could be radial excitations or non-resonant combinations of undetermined quantum numbers.   Thus the DELPHI data taken by themselves are 
{\em not necessarily} in 
conflict with theoretical expectations. 

However the plot thickens in several respects: 
\begin{itemize}
\item 
BELLE has presented an analysis this summer of $\bar B\to l \bar \nu D/D^* \pi$ \cite{BELLESL}, which appears to be in conflict with previous findings. Reconstructing one $B$ completely in 
$\Upsilon (4S) \to B \bar B$, they analyze the decays of the other beauty meson and obtain: 
\bea 
{\rm BR}(B^- \to l^- \bar \nu D\pi ) &=& (0.81 \pm 0.18) \% 
\label{BELLESL1}\\
{\rm BR}(B^- \to l^-  \bar \nu D^*\pi ) &=& (1.00 \pm 0.22) \% 
\label{BELLESL2}\\
{\rm BR}(\bar B_d \to l^-  \bar \nu D\pi ) &=& (0.49 \pm 0.13) \% 
\label{BELLESL3}\\
{\rm BR}(\bar B_d \to l^- \bar \nu D^*\pi ) &=& (0.97 \pm 0.22) \%
\label{BELLESL4}
\eea
BELLE's separation of final states with $D$ and $D^*$ is of significant value, 
since it provides information on the relative weight of `3/2' and `1/2' production. 

Combining the two classes of final states they arrive at: 
\bea 
{\rm BR}(B^- \to l^- \bar \nu D^{(*)}\pi ) &=& (1.81 \pm 0.20 \pm 0.20) \% \\
{\rm BR}(\bar B_d \to l^- \bar \nu D^{(*)}\pi ) &=& (1.47 \pm 0.20 \pm 0.17) \%
\eea
leaving room for a large $D^{(*)}\pi\pi$ component of $\sim (1.3 \pm 0.4)\%$, whereas 
previous studies have obtained 90\% C.L. upper limits ranging from $0.35$ to $1.3$ \%. 

BELLE's numbers are actually quite consistent with the theoretical predictions the BT model yields for `3/2' P wave production; yet BELLE has not determined the quantum numbers of their  hadronic final states. 

\item 
The BT model predicts for $D\pi$ and $D^*\pi$ production: 
\bea 
{\rm BR}(B^- \to l^- \bar \nu D\pi ) &=& {\rm BR}(\bar B_d \to l^- \bar \nu D\pi ) = 
0.51  \% \\
{\rm BR}(B^- \to l^- \bar \nu D^*\pi ) &=& {\rm BR}(\bar B_d \to l^-  \bar \nu D^*\pi )=
0.65 \% 
\eea
in qualitative agreement with BELLE's numbers. 

\item 
In the BPS approximation \cite{BPS} one has $\tau ^{(n)}_{1/2}=0$. Assuming that the 
sum rule of Eq.(\ref{SR1}) saturates already with the $n=0$ state, one obtains 
$\tau ^{(0)}_{3/2}=\frac{1}{2}$ leading to 
\bea 
{\rm BR}(B^- \to l^- \bar \nu D\pi ) &=& {\rm BR}(\bar B_d \to l^- \bar \nu D\pi ) = 
0.39  \% \\
{\rm BR}(B^- \to l^- \bar \nu D^*\pi ) &=& {\rm BR}(\bar B_d \to l^-  \bar \nu D^*\pi )=
0.50 \%  \; . 
\eea
\item 
Using the experimental numbers stated in Eqs.(\ref{DELPHI3/2}, \ref{DELPHI1/2}) and assuming 
that the "1" and "0" state decay 100 \% into $D^*\pi$ and $D\pi$, respectively, one arrives at 
\bea 
{\rm BR}(B^- \to l^- \bar \nu D\pi ) &=& {\rm BR}(\bar B_d \to l^- \bar \nu D\pi ) \sim  
(0.9 \pm 0.7)  \% \\
{\rm BR}(B^- \to l^- \bar \nu D^*\pi ) &=& {\rm BR}(\bar B_d \to l^-  \bar \nu D^*\pi ) \sim 
(1.9 \pm 0.4) \%  \; . 
\eea
for a total of 
\beq 
{\rm BR}(B^- \to l^- \bar \nu D^{(*)}\pi ) = {\rm BR}(\bar B_d \to l^-  \bar \nu D^{(*)}\pi ) \sim 
2.8 \%
\eeq
One should note that the qualitative trend is the same as with BELLE's findings, 
Eqs.(\ref{BELLESL1} - \ref{BELLESL4}) -- namely that $D^*\pi$ final states dominate over 
$D\pi$ ones -- yet the total $D^{(*)}\pi$ rate exceeds that reported by BELLE and predicted by 
the BT model by about 1 percentage point. This is of course just a rephrasing of the `1/2 vs. 3/2' puzzle.

\item 
Theoretically both in the framework of the BT model and purely of the OPE treatment one does {\em not}   expect the $D$, $D^*$ and four P wave $D^{**}$ states to {\em completely} saturate the inclusive width, 
as mentioned above. 
This can be seen most explicitly in the BT quark model calculation. Likewise in the 
OPE treatment one expects a broad mass distribution at the higher end. 
As already mentioned at the end of Sect.\ref{OPETR}, 
already ${\cal O}(\alpha_S)$ perturbative corrections not included so far will 
generate a smooth 
tail in the mass distribution of the hadronic final state towards the upper end of the mass spectrum. 
Of course those will not be clear resonances, but a broad and continuous distribution of masses -- in close analogy what happens in $e^+e^- \to had.$ when one goes more and more above charm threshold. 

There is no sign of such high mass combinations in the CDF data \cite{CDFMOM} 
and no obvious one in the DELPHI data beyond the tail of their $D^{**}$ states. More specifically  one finds  
that about 6.4\% and 18.3\% of all $D^{**}$ states have masses between 2.6 and 3.3 GeV for 
the CDF and DELPHI data, respectively, which drop to 3.2 \% and 7.8\% for the mass range 
2.8 to 3.3 GeV and 0.3 \% and 3.1 \% for 3.0 to 3.3 GeV. 

On the other hand, CDF seems to see more events below the $D^{3/2}$ peaks.  
Such low mass $D^{(*)}\pi$ events could be due to higher mass states decaying into 
$D^{(*)}\pi\pi$. CDF has not incorporated this scenario into their analysis, since previous measurements showed no evidence for such decays. 

\item 
One would conjecture that if the observed mass spectrum indeed differs significantly from theoretical expectations -- in its center of gravity as well as its spread --, then the measured hadronic mass moments should not follow theoretical predictions -- 
yet they do \cite{FLAECHER,BABARVCB,BATTAG,DELPHI,CDFMOM}. 
\end{itemize}

{\em In summary:} ALEPH, DELPHI and D0 agree in finding a rate of about 0.8 \% of $\Gamma _B$ 
for the production of the two {\em narrow} $D^{**}$ states combined. This value is quite consistent 
with theoretical expectations for the $D^{3/2}$ rates. BELLE's data also fit naturally into this picture. 
The problem arises in the production of the {\em broad} $D^{**}$ states: The rate found by 
ALEPH and DELPHI suffice to saturate $\Gamma _{SL}(B)$, yet exceed the predictions for 
$\bar B \to l \bar \nu D^{1/2}$ by about an order of magnitude. BELLE's numbers on the other 
hand agree reasonably well with predictions, yet fall short of saturating $\Gamma _{SL}(B)$.

%%%%%%%%%%
\section{Conclusions and call for action}
\label{ACTION}
%%%%%%%%%%%%%

The theoretical predictions on $\bar B \to l \bar \nu X_c$ described here have a solid foundation. The OPE treatment is genuinely based on QCD, and while the BT description invokes a model, it 
implements QCD dynamics for heavy flavour hadrons to a remarkable degree. Their prediction therefore deserve to be taken seriously and not discarded at the first sign of phenomenological 
trouble. Preliminary lattice studies show no significant enhancement of 
`1/2' production. The numbers we have given for the theoretical expectations should be taken with 
quite a few grains of salt. Yet the predicted pattern that the abundance of `3/2' P wave resonances 
dominates over that for `1/2' states in semileptonic $B$ decays is a sturdy one. 

The $B_d$ and $B_u$ semileptonic widths have been well measured. Most if not even all of it 
has been identified in $\bar B \to l \bar \nu D/D^* +(0,1)\pi$. The next important steps are 
\begin{itemize}
\item 
to clarify the size, mass distribution and quantum numbers of $\bar B \to l \bar \nu [D/D^*\pi]_{\rm broad}$ 
and 
\item 
to search for $\bar B \to l \bar \nu D/D^* + 2 \pi$ with even higher sensitivity. 
\item 
The data should be presented {\em separately} for $\bar B\to l \bar \nu D+\pi$'s and 
$\bar B\to l \bar \nu D^*+\pi$'s, 
since it provides more theoretical diagnostics. 

\end{itemize}
These are challenging experimental tasks, yet highly rewarding ones as well: 
\begin{itemize}
\item 
They probe our theoretical control over QCD's nonperturbative dynamics in novel and sensitive ways. 
This is an area where different theoretical technologies -- the OPE, quark models and 
lattice QCD -- are making closer and closer contact. 

The lessons to be learnt will be very significant ones, no matter what the eventual experimental verdict will be: 
\begin{itemize}
\item 
A confirmation of the OPE expectations and even the more specific BT predictions would reveal an even higher degree of theoretical control over nonperturbative QCD dynamics than has been 
shown through $\Gamma (\bar B \to l \bar \nu X_c)$. 
\item 
Otherwise we could infer that formally nonleading $1/m_Q$ corrections are highly significant numerically. Such an insight would be surprising -- yet important as well. In particular 
it would provide a highly nontrivial challenge to lattice QCD. Meeting this challenge 
successfully would provide 
lattice QCD with significantly enhanced validation. 
\end{itemize}
\item 
On the more pragmatic side one should note that understanding the hadronic final state in 
semileptonic $B$ decays is of crucial importance, when measuring moments of the hadronic 
recoil mass spectrum in $\bar B \to l \bar \nu X_c$. From those moments -- the average, variance etc. -- 
one extracts the values of the heavy quark parameters $m_b$, $m_c$, $\mu _{\pi}^2$ etc. with high accuracy for their intrinsic interest and as input to 
determinations of $|V(cb)|$ and $|V(ub)|$. 
\end{itemize}
The call for further action is directed to theorists as well: 
\begin{itemize}
\item 
The impact of perturbative QCD corrections on the OPE description of higher states in 
$\bar B \to l \bar \nu X_c$ should be evaluated quantitatively. 
\item 
In the BT model one can -- and should -- compute the production rates for the higher orbital and radial 
excitations in semileptonic $B$ decays. 
\item 
The strong decays $D^{**} \to D/D^* + \pi \pi$ should be estimated using heavy quark symmetry 
arguments augmented by quark model considerations. 
\item 
The BT model predictions were obtained in the heavy quark limit. Corrections to this limit 
could be quite important as suggested 
in Ref.\cite{lig}, and they could significantly change the relative weight of 
$\tau ^{(n)}_{1/2}$ and $\tau ^{(n)}_{3/2}$. Calculating or at least constraining those corrections  would be a most worthwhile undertaking -- 
alas it requires some new ideas. A priori one can conceive of different ways of extending the 
BT description to include finite mass effects, yet they are unlikely to be equivalent. The foundations 
for a promising way have been laid in Ref.\cite{OPTICAL}. 
% For although some form factors 
%are known at zero recoil, many unknowns remain.  
\item 
Lattice QCD studies of `1/2' and `3/2' production in semileptonic $B$ decays has to be 
pursued with vigour. Such studies could turn out to be veritable `gold mines' as far as validation is concerned.  One can evaluate the spectrum of the higher radial and orbital excitations $D^{**}$, for which some encouraging results have already been obtained \cite{UKQCD}. Lattice 
calculations at finite values of $m_c$ should be performed, which would teach us about 
$1/m_c$ corrections. 

\end{itemize}

In other words: Since there is a lot to be done, we better get started! 

\vspace{0.5cm}

\noindent
{\bf Acknowledgments:} This work was supported  by the NSF under grant number PHY-0355098 
and by the EC contract HPRN-CT-2002-00311 (EURIDICE).

%%%%%%%%%%%%

\end{document}